\documentclass[10pt, conference, twocolumn]{IEEEtran}

\IEEEoverridecommandlockouts

\usepackage{amsmath,amssymb,amsfonts}
\usepackage{graphicx}
\usepackage{textcomp}

\usepackage{optidef}
\usepackage[nocompress]{cite}

\usepackage{rotating}
\usepackage{multirow}
\usepackage{array}
\usepackage{xcolor}
\usepackage{url}
\usepackage{pgfplots}
\usepackage{verbatim}
\usepackage{tikz}
\usepackage{algorithm}
\usepackage{algorithmic}
\usepackage{breqn}
\usepackage{amsmath}
\usepackage{aurical}
\usepackage[T1]{fontenc}

\usepackage[utf8]{inputenc}
\usepackage{nomencl}
\usepackage[normalem]{ulem}
\usepackage{soul}
\makenomenclature

\usepackage{subfig}

\newcommand{\tab}{\hspace{3pt}}
\newcommand{\tabmore}{\hspace{10pt}}

\newlength\myindent
\usepackage{mathtools}

\usepackage{color, colortbl}
\definecolor{LightCyan}{rgb}{0.88,1,1}

\def\BibTeX{{\rm B\kern-.05em{\sc i\kern-.025em b}\kern-.08em T\kern-.1667em\lower.7ex\hbox{E}\kern-.125emX}}

\usetikzlibrary{mindmap,shadows,backgrounds}
\graphicspath{{figures/}}
\pgfplotsset{compat=1.14}

\def\BibTeX{{\rm B\kern-.05em{\sc i\kern-.025em b}\kern-.08em
    T\kern-.1667em\lower.7ex\hbox{E}\kern-.125emX}}

\ifCLASSINFOpdf

\else

\fi

\begin{document}

\title{Radio Resource and Beam Management in 5G mmWave Using Clustering and Deep Reinforcement Learning}

\author{\IEEEauthorblockN{Medhat Elsayed, \IEEEmembership{Student Member, IEEE}, and Melike Erol-Kantarci, \IEEEmembership{Senior Member, IEEE}} \\
\IEEEauthorblockA{School of Electrical Engineering and Computer Science \\ University of Ottawa \\ Ottawa, Canada, \\
Email: \{melsa034, melike.erolkantarci\}@uottawa.ca}
}

\maketitle

\begin{abstract}
To optimally cover users in millimeter-Wave (mmWave) networks, clustering is needed to identify the number and direction of beams. The mobility of users motivates the need for an online clustering scheme to maintain up-to-date beams towards those clusters. Furthermore, mobility of users leads to varying patterns of clusters (i.e., users move from the coverage of one beam to another), causing dynamic traffic load per beam. As such, efficient radio resource allocation and beam management is needed to address the dynamicity that arises from mobility of users and their traffic. In this paper, we consider the coexistence of Ultra-Reliable Low-Latency Communication (URLLC) and enhanced Mobile BroadBand (eMBB) users in 5G mmWave networks and propose a Quality-of-Service (QoS) aware clustering and resource allocation scheme. Specifically, Density-Based Spatial Clustering of Applications with Noise (DBSCAN) is used for online clustering of users and the selection of the number of beams. In addition, Long Short Term Memory (LSTM)-based Deep Reinforcement Learning (DRL) scheme is used for resource block allocation. The performance of the proposed scheme is compared to a baseline that uses K-means and priority-based proportional fairness for clustering and resource allocation, respectively. Our simulation results show that the proposed scheme outperforms the baseline algorithm in terms of latency, reliability, and rate of URLLC users as well as rate of eMBB users.
\end{abstract}

\IEEEpeerreviewmaketitle

\section{Introduction}\label{sec:intro}
With the unprecedented growth of mobile data traffic stemming from a growing use of data-hungry applications, next-generation wireless networks need to adopt a paradigm shift in the way the resources are managed. Millimeter Wave (mmWave) technology is promising large and underutilized spectrum between $30$ and $300$ GHz, which addresses the well-known spectrum scarcity problem of the sub-$6$ GHz band \cite{6824752}. However, mmWave suffers from high propagation losses that hinder its coverage range. One approach to combat such losses is to use directional communication where beamforming is used to reshape the pattern of propagation in the direction of the user. 

Despite the performance gains that beamforming alongside mmWave bring about, many challenges exist. The distribution of users and traffic can vary rapidly within a short period of time \cite{8968715}. The fifth generation (5G) standard introduced three service categories: Ultra-Reliable Low-Latency (URLLC), enhanced Mobile Broad Band (eMBB), and massive machine-type communication (mMTC) \cite{ITUR}. In addition, wireless networks beyond 5G  and 6G are expected to serve applications with more heterogeneity and tight Quality-of-Service (QoS) requirements \cite{DBLP}. Furthermore, an added layer of complexity arises due to mobility of users. With such network dynamicity, beam management and radio resource allocation becomes more challenging. This, first, calls for an intelligent beam management algorithm that captures Qos and mobility of users. Second, an intelligent radio resource allocation is needed to actively consider load variations across the formed beams. 

In this paper, we consider a heterogeneous mmWave network that employs beamforming for serving URLLC and eMBB users. Since users are mobile, an online clustering is sought to cluster users that can be served by a single beam. In addition, due to the fact that load per beam changes as users move among clusters, Resource Block (RB) allocation is needed to efficiently allocate resources among users within the same beam. For this purpose, we propose a QoS-aware clustering and resource block allocation technique for mmWave networks. In particular, we propose a DBSCAN-based algorithm for user clustering and managing beams, in addition to a Long Short-Term Memory (LSTM)-based deep reinforcement learning for RB allocation. We call our algorithm as Deep Q-learning with DBSCAN (DQLD). Furthermore, we compare the proposed algorithm to a baseline algorithm that employs K-means clustering instead of DBSCAN and Priority-based Proportional Fairness (KPPF) instead of DRL for resource allocation. Simulation results reveal that DQLD outperforms KPPF in latency, reliability, and rate of URLLC users as well as rate of eMBB users. 

This paper is structured as follows. Section \ref{sec:relWork} presents the related work. Section \ref{sec:sysModel} introduces the system model and the problem formulation. In section \ref{sec:alg}, the proposed algorithm for improving network rate is discussed. Section \ref{sec:perfEval} presents the simulation setup, baseline algorithm and performance results. Finally, section \ref{sec:conclusion} concludes the paper. 
\section{Related Work}\label{sec:relWork}
Machine learning, and specifically deep reinforcement learning, is gaining more popularity in the applications of wireless networks \cite{8714026}. An unsupervised machine learning method was proposed in \cite{9083909} for automatic identification of the optimal number of clusters in mmWwave Non-Orthogonal Multiple Access (NOMA). The proposed clustering is an agglomerative hierarchical clustering that infers the number of clusters for maximum sum rate in the network. In \cite{8536429}, authors consider a mmWave network that comprises femto access points and femto users and propose a clustering algorithm for improving network rate. Furthermore, the authors propose a joint user-cell association and resource allocation to further improve the performance. The solutions are formulated as optimization problems and solved with some simplifications. The work in \cite{8938771} addresses the joint design of beamforming, power, and interference coordination in a mmWave network. The authors formulate a non-convex optimization problem to maximize Signal-to-Interference plus Noise Ratio (SINR) and solve it using deep reinforcement learning. In \cite{8454272}, the authors consider a mmWave NOMA system and aim to maximize the sum rate of the network using clustering and Power Domain NOMA (PD-NOMA). In particular, clustering relies on the fact that adjacent users experience similar channel. Furthermore, authors derive closed form expressions for optimal NOMA power allocation. 

Unlike previous works, we address the dynamicity in a mmWave network using joint online clustering and resource block allocation. The proposed algorithm aims to address network conditions as well as heterogeneous traffic captured through user-specific QoS requirements. As such, we propose an LSTM-based deep reinforcement learning algorithm for resource allocation. 

In \cite{icc2020}, we proposed a Q-learning algorithm for joint user-cell association and inter-beam power allocation. The proposed algorithm focused on improving the network sum rate when applying NOMA in mmWave networks. This paper differs from our previous work in two aspects. First, we employ orthogonal multiple access instead of NOMA and propose deep reinforcement learning algorithm for radio resource allocation. In addition, here, we employ DBSCAN for user clustering and to the best of our knowledge, this is the first time DBSCAN and DRL is used jointly for resource management. 
\section{System Model}\label{sec:sysModel}
\subsection{Network Model}
\begin{figure}
    \centering
    \includegraphics[scale=0.32]{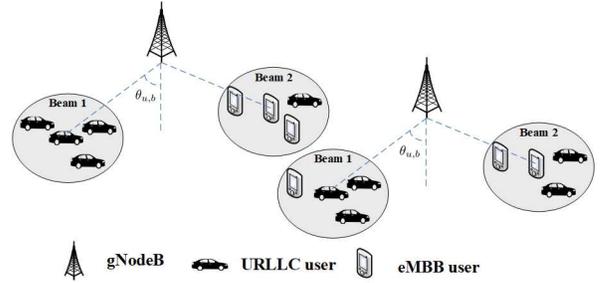}
    \caption{System model of mmWave network.}
    \label{fig:sysModel}
\end{figure}

Consider a mmWave network with $g \in$ {\Fontauri\bfseries G} of 5G-NodeBs (gNBs), where each gNB covers $U \in$ {\Fontauri\bfseries U} single-antenna users. Users are partitioned into different clusters, where each cluster is served by a single beam denoted by $b \in$ {\Fontauri\bfseries B} as shown in Fig. \ref{fig:sysModel}. We consider two types of users with different QoS: URLLC and eMBB users. In particular, URLLC users require a low latency and high reliability communication, whereas eMBB users require high rate communication. Let {\Fontauri\bfseries U}$_b$ be the set of users covered by $b^{th}$ beam and the communication between beams and their associated users follows 5G-NR release 15 \cite{TS38211}. Furthermore, beams use Orthogonal Frequency Division Multiple Access (OFDMA) to allocate orthogonal resources to their users, hence intra-beam interference can be omitted. The bandwidth, $\omega_b$, of $b^{th}$ beam is subdivided into a number of Resource Blocks (RBs), where a RB consists of $12$ subcarriers. Furthermore, contiguous RBs are grouped to form a RBG. Let $k \in$ {\Fontauri\bfseries K} denote a RBG and the bandwidth of a RBG is denoted by $\omega_{k,b}$. The duration of RB (or RBG) can span multiple OFDM symbols in time, which is denoted as Transmission Time Interval (TTI). Hence, the minimum resource allocation is considered to be one RBG, where we select $2$ OFDM symbols as the length of a TTI to encourage low latency for URLLC users \cite{8647289, 8361404}.   

The initial positions of users follow a Poisson Cluster Process (PCP), in which heads of clusters are uniformly distributed and users within each cluster are uniformly distributed within the radius of the cluster. In addition, mobility of users follow random waypoint mobility model. The traffic of users follows Poisson distribution with $\lambda$ inter-arrival time and a fixed packet size of $32$ bytes. As such, users tend to leave their clusters and join new ones as time proceeds.
The mmWave channel can be modeled using a single Line-of-Sight (LoS) path model, where the gain of the LoS path is larger than the gain of non-Line-of-Sight (nLoS) paths \cite{8454272}. As such, the channel vector, $\boldsymbol{h}_{k,u,b} \in \mathbb{C}^{M \times 1}$, between $b^{th}$ beam and $u^{th}$ user on $k^{th}$ RBG is represented as follows:
\begin{equation}
    \boldsymbol{h}_{k,u,b}= \boldsymbol{v}(\theta_{u,b}) \frac{\alpha_{k,u,b}}{\sqrt{M}(1 + d_{u,b}^n)},
\end{equation}
where $\alpha_{k,u,b} \in \mathbb{C}N(0, \sigma^2)$ is the complex gain, $M$ is the number of paths, $d_{u,b}$ is the Euclidean distance, and $n$ is the pathloss exponent. Note that we removed the gNB's index to keep the formulation readable. In addition, $\boldsymbol{v}(\theta_{u,b})$ is the steering vector, which can be represented as follows:
\begin{equation}
    \boldsymbol{v}(\theta_{u,b}) = [1, e^{-j 2 \pi \frac{L}{\lambda} \sin(\theta_{u,b})}, ..., e^{-j 2 \pi (N-1) \frac{L}{\lambda} \sin(\theta_{u,b})}]^T,
\end{equation}
where $L$ is the gNB's antenna spacing, $N$ is the number of antenna's elements, $\lambda$ is the wavelength, $\theta_{u,b}$ is the Angle of Departure (AoD).

\subsection{QoS Requirements}
The proposed scheme aims at addressing the QoS differences among the users in the network. In particular, URLLC users need to maintain high reliability and low latency communication links, whereas eMBB users need to achieve high rate. 

\subsubsection{Rate of eMBB users}
The sum rate of eMBB users per gNB is formulated as follows:
\begin{equation}
    C = \sum\limits_{b \in \text{\Fontauri\bfseries B}} \sum\limits_{u \in \text{\Fontauri\bfseries U}_b^e} \sum\limits_{k \in \text{\Fontauri\bfseries K}_b} \delta_{k,u,b} \omega_{k,b} \log_2(1 + \Gamma_{k,u,b}),
    \label{eq:sumRate}
\end{equation}
where $\delta_{k,u,b}$ is a RBG allocation indicator, $\omega_{k,b}$ is the size of RBG in Hz, $\text{\Fontauri\bfseries U}_b^e$ is the set of eMBB users that belong to $b^{th}$ beam, and $\Gamma_{k,u,b}$ is the SINR of $(k,u,b)^{th}$ link, which can be expressed as
\begin{equation}
    \Gamma_{k,u,b} = \frac{p_{k,b} |\boldsymbol{h}_{k,u,b}^H \boldsymbol{w}_{k,b}|^2}{\sigma^2 + \sum\limits_{b' \neq k} p_{k,b'} |\boldsymbol{h}_{k,u',b'}^H \boldsymbol{w}_{k,b'}|^2},
    \label{eq:sinr}
\end{equation}
where $p_{k,b}$ and $\boldsymbol{w}_{k,b}$ denote the power and beamforming vector of $k^{th}$ RBG of $b^{th}$ beam. $p_{k,b'}$ and $\boldsymbol{w}_{k,b'}$ denote the power and beamforming vector of $k^{th}$ RBG of ${b'}^{th}$ interfering beam. $\sigma^2$ represents receiver's noise variance. 

\subsubsection{Latency and reliability of URLLC users}
Latency of URLLC users is formulated as follows:
\begin{equation}
    D_{u,b} = D_{u,b}^{tx} + D_{u,b}^{q} + D_{u,b}^{harq},
    \label{eq:latencyUrllc}
\end{equation}
where $D_{u,b}^{tx}$ is the transmission latency, $D_{u,b}^{q}$ is the queuing latency (i.e., latency of packet pending in the transmission buffer), and $D_{u,b}^{harq}$ is the Hybrid Automatic Repeat Request (HARQ) re-transmission latency of $u^{th}$ user on $b^{th}$ beam. In line
with \cite{7962790}, we assume $D_{u,b}^{harq} = 4 TTI$. In particular, queuing and re-transmission latencies constitute the dominant factors in eq. (\ref{eq:latencyUrllc}) \cite{8517001}. Furthermore, the queuing latency is a direct outcome of the decision of the scheduler (i.e., queuing and scheduling latencies are identical). As such, in order to achieve low latency for URLLC users, the scheduler has to immediately allocate resources to URLLC traffic once it arrives. Furthermore, the number of re-transmissions has to be limited, where in this work, we assume $1$ HARQ re-transmission. 

Limiting the number of re-transmissions, however, may impact the reliability of URLLC users. To maintain high reliability, link adaptation is performed, where users periodically report SINR measurements to the gNB in the form of Channel Quality Indicator (CQI) values. CQI indicates the quality (i.e., SINR) of the link with the associated beam. In turn, the gNB accounts for those measurements in the scheduling policy. In the following section, we show how the proposed DRL addresses both latency and reliability of URLLC users.
\section{Deep Q-learning with DBSCAN (DQLD)}\label{sec:alg}
In order to maintain high QoS of URLLC and eMBB in the midst of changing network conditions, the proposed algorithm considers an online clustering (for the purpose of beam management) and a machine-learning based resource allocation. Online clustering is used to cluster users that are adjacent to each other and can be covered by a single beam. In addition, the online clustering algorithm aims to find the optimal number of beams for coverage. On the other hand, for resource block allocation we use deep Q-learning. The joint beam and resource management scheme is Deep Q-learning with DBSCAN (DQLD). DBSCAN is used for online clustering and deep Q-learning is used for resource block allocation. 
An online algorithm is needed to maintain efficient coverage of mobile users. In this work, we adopt DBSCAN for user clustering and selection of number of beams due to its advantages over other clustering techniques. \cite{3001507}. DBSCAN does not require a predefined number of clusters. Instead, the algorithm identifies users that can belong to a cluster from sparse users and returns the number and structure of clusters. In addition, DBSCAN has low complexity and easy implementation.  

Note that, frequent online clustering can lead to a challenging resource allocation problem. In particular, performing the clustering very often leads to frequent changes in the structure and number of beams. As such, resource allocation has to deal with a very dynamic environment. Furthermore, clustering might be needed only whenever the beams are not efficient enough to cover users (i.e., users have changed their positions and tend to belong to new clusters). Therefore, determining the frequency of clustering is important. We choose to perform clustering only when the average SINR of a beam drops under a predefined threshold. 

Clustering returns a set of beams to cover network users. Within each beam, we perform resource block allocation using an LSTM-based deep Q-learning technique, namely DQL. The tuples of DQL is defined as follows:
\subsubsection{Agents} DQL is a multi-agent distributed algorithm that is performed independently by each gNB (i.e., each gNB is a standalone agent). Each gNB performs DQL to allocate RBGs within each of its beams. 
\subsubsection{Actions} The actions are defined as the RBGs allocated to users per beam as
\begin{equation}
    a_{k,b} = \{u_{k,b}\},
\end{equation}
where $a_{k,b}$ denotes the action of $k^{th}$ RBG of $b^{th}$ beam, and $u_{k,b}$ is the user index. 
\subsubsection{States} We design the states in a way that captures the level of inter-beam interference. In particular, states are defined in terms of the CQI feedback measured at the user. Therefore, state of $k^{th}$ RBG of $b^{th}$ beam is defined as
\begin{equation}
    s_{k,b} = \{q_{k,b}\},
    \label{eq:state}
\end{equation}
where $q_{k,b}$ is the CQI of $k^{th}$ RBG of $b^{th}$ beam. 
\subsubsection{Reward} The reward function is designed to account for different users' classes (i.e., URLLC and eMBB). In particular, URLLC users require tight latency and reliability, whereas eMBB users require high throughput. Therefore, the reward function is defined as
\begin{equation}
    r_{k,b} = 
    \begin{cases}
     \text{sigm}(r_{k,b}^{(mbb)} \tab r_{k,b}^{(llc)}), \tabmore \tab \tab C(u) = 1, \\
    \text{sigm}(r_{k,b}^{(mbb)}), \tabmore \tabmore \tabmore \tabmore  C(u) = 2,
    \end{cases}
    \label{eq:reward}
\end{equation}
where $\text{sigm}(x)$ denotes a sigmoid function defined as
\begin{equation}
    \text{sigm}(x) = \frac{1}{1 + e^{-x}}.
    \label{eq:sigmoid}
\end{equation}
In eq. (\ref{eq:reward}), $C(u)$ presents the Quality Class Indicator (QCI) of $u^{th}$ user, where $C(u) = 1$ denotes URLLC users, while $C(u) = 2$ denotes eMBB users. $r_{k,b}^{(mbb)}$ and $r_{k,b}^{(llc)}$ are the reward functions of eMBB and URLLC users, respectively, which are defined as follows:
\begin{equation}
    r_{k,b}^{(llc)} = \frac{D^{QoS}}{D_{k,b}(u)},
    \label{eq:urllcReward}
\end{equation}
\begin{equation}
    r_{k,b}^{(mbb)} = \frac{\Gamma_{k,b}}{\Gamma^{QoS}},
    \label{eq:embbReward}
\end{equation}
where $D_{k,b}(u)$ is the queuing latency due to allocating $k^{th}$ RBG of $b^{th}$ beam to $u^{th}$ user, $D^{QoS}$ is the QoS requirement of latency, and $\Gamma^{QoS}$ is the QoS requirement of SINR. It is worth mentioning that gNB has knowledge of QCI of its radio bearers and queuing latency of its users. As such, when traffic on link $(k,b)$ belongs to a URLLC user, the reward constitutes a combination of reliability and queuing latency. Indeed queuing latency dominates the total latency of downlink transmission \cite{8911618}. On the other hand, when traffic on link $(k,b)$ belongs to a eMBB user, the reward constitutes reliability, which translates to higher transmission throughput (i.e., improving SINR enables allocation of higher modulation and coding scheme and higher transport block size). Finally, the sigmoid function is used to keep the reward in the interval $[0,1]$.

\begin{figure}
    \centering
    \includegraphics[scale=0.3]{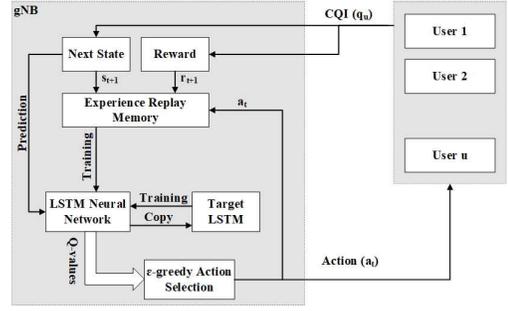}
    \caption{Conceptual diagram of LSTM-based Deep Q-learning.}
    \label{fig:lstmDrl}
\end{figure}

Fig. \ref{fig:lstmDrl} presents a conceptual diagram of the LSTM-based DQL approach. It is worth mentioning that gNB has a separate DQL entity for each beam it forms. For each beam, the DQL works as follows. The gNB computes the next state and the reward as in (\ref{eq:state}) and (\ref{eq:reward}), respectively, from the CQI and SINR feedback received from its users. The experience, $\{s_t, a_t, r_{t+1}, s_{t+1}\}$, is then stored in the experience replay memory to be used later for training the LSTM neural network, where $s_t, a_t, r_{t+1},$ and $s_{t+1}$ are the state, action at $t^{th}$ time step, reward, and next state at $(t+1)^{th}$ time step. Afterwards, LSTM is used to predict the Q-values of all actions of the next state (i.e., $Q(s_{t+1}, a')$. Finally, the Q-values is fed to the $\epsilon$-greedy algorithm for next action selection. The $\epsilon$-greedy algorithm selects either a random action with probability $(\epsilon)$ or an action that follows the greedy policy with probability $(1-\epsilon)$. 

To maintain low complexity, the training of LSTM is performed every $T$ TTIs. In particular, a batch of experience samples is drawn randomly from the experience replay memory. The batch is fed to the target LSTM in order to compute a sequence of reference responses. These responses constitute the labels used to train the main LSTM network. In addition, the target LSTM is initially loaded with the weights of the main network. However, the update of the target LSTM's weights are done every $C$ TTIs to maintain stability \cite{playingAtari}. 
\section{Performance Evaluation}\label{sec:perfEval}

\subsection{Simulation Setup}
We perform simulation using a discrete event simulation based on 5G Matlab Toolbox. Table \ref{tab:simulationSettings} presents network and simulator, and DQLD algorithm settings. The network is composed of two gNBs with $300$ m inter-gNBs distance. Initial positions of users follow a PCP with $2$ clusters, where each cluster has a radius of $20$ m. The performance of the algorithms is tested under different traffic loads (i.e., \{$0.5, 1, 1.5,2\}$ Mbps per gNB). The DQLD algorithm consists of $15$ states (i.e., corresponding to number of CQIs) and $24$ actions (i.e., actions correspond to total number of users per gNB).  
\begin{table*}[ht]
    \centering
    \caption{Simulation Settings}
    \begin{tabular}{|l|l|}
         \hline
         \textbf{\underline{5G PHY configuration}} &  \\
         Bandwidth & $20$ MHz \\
         Carrier frequency & $30$ GHz \cite{8782638}\\
         Subcarrier spacing & $15$ KHz \\
         Subcarriers per resource block & $12$ \\
         TTI size & $2$ OFDM symbols ($0.1429$ msec) \\
         Max transmission power & $28$ dBm \\
         URLLC target BLER & $1\%$\\
         eMBB target BLER & $10\%$\\
         \textbf{\underline{HARQ}} & \\
         Type & Asynchronous HARQ \\
         Round trip delay & $4$ TTIs \\
         Number of processes & $6$ \\
         Max. number of re-transmission & $1$ \\
         \textbf{\underline{Network model}} & \\
         Initial positions & Poisson Cluster Process (PCP) \\
         Mobility & Random waypoint \\ 
         number of URLLC per cluster & $2$ \\
         number of eMBB per cluster & $2$ \\
         Number of clusters & $3$ \\
         Radius of cluster & $20$ m \\
         Number of gNBs & $2$ \\
         Radius of cell & $150$ m\\
         Inter-site distance & $300$ m \cite{8782638} \\
         \hline
    \end{tabular}
    \begin{tabular}{|l|l|}
         \hline
         \textbf{\underline{Traffic}} & \\
         Distribution & Poisson \\
         Packet size & $32$ Bytes \\
         \textbf{\underline{Q-learning}} & \\
         Learning rate $(\alpha)$ & $0.5$ \\
         Discount factor $(\gamma)$ & $0.9$ \\
         Exploration probability $(\epsilon)$ & $0.1$ \\
         $D^{QoS}$ & $1$ msec\\
         $\Gamma^{QoS}$ & $15$ dB \cite{6777898} \\
         \textbf{\underline{LSTM}} & \\
         Size of input layer & $1$ \\
         Number of hidden units & $20$ \\
         Size of output layer & $24$\\
         Size of mini-batch & $20$ \\
         Size of replay memory & $60$ \\
         Training Interval $(T)$ & $60$ \\
         Copy Interval $(C)$ & $120$ \\
         \textbf{\underline{DBSCAN}} & \\
         $minPts$ & $5$ \\
         $eps$ & $30$ \\
         \textbf{\underline{Simulation parameters}} & \\
         Simulation time & $1.5$ second \\
         Number of runs & $10$ \\
         Confidence interval & $95\%$ \\
         \hline
    \end{tabular}
    \label{tab:simulationSettings}
\end{table*}

\subsection{Baseline Algorithm}
For fair comparison, we use a baseline algorithm that works in a similar fashion to the proposed algorithm and has been used in the literature before. In the baseline, K-means is used to perform online clustering  \cite{8454272}, whereas priority-based proportional fairness is used for resource block allocation as proposed in \cite{8361404}. In \cite{8454272}, clustering using K-means is performed based on channel properties at the user side, i.e., users that are close in proximity are more likely to experience similar channels. Furthermore, in \cite{8454272}, resource block allocation is performed using a hard QoS-aware criteria. In particular, RBGs are given to URLLC users with pending data transmission first, then the remaining RBGs are allocated to eMBB users. Within each user class, RBGs are distributed according to proportional fairness criteria as
\begin{equation}
    u^* = \arg \max \frac{C_{k,u,b}}{\overline{C}_{k,u,b}},
\end{equation}
where $u^*$ is the selected user to be allocated $k^{th}$ RBG.

\subsection{Performance Results} 
\begin{figure}[t]
    \centering
    \includegraphics[scale=0.27]{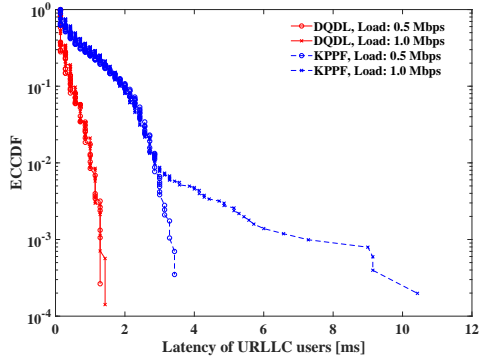}
    \caption{Latency of URLLC users versus total URLLC offered load ($[0.5, 1]$ Mbps).}
    \label{fig:urllcLatency01}
\end{figure}

\begin{figure}[t]
    \centering
    \includegraphics[scale=0.27]{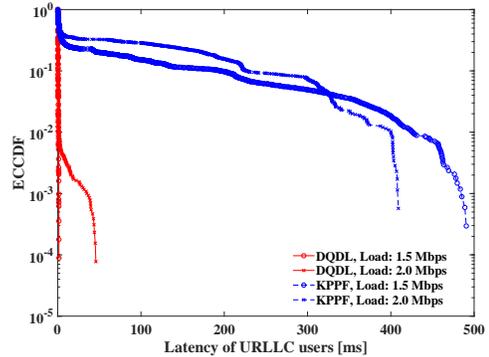}
    \caption{Latency of URLLC users versus total URLLC offered load ($[1.5, 2]$ Mbps).}
    \label{fig:urllcLatency02}
\end{figure}

In this subsection, we present the simulation results of the proposed DQLD scheme and compare it to the baseline KPPF algorithm. The performance is assessed in terms of URLLC and eMBB QoS requirements. Fig. \ref{fig:urllcLatency01} and Fig. \ref{fig:urllcLatency02} present the Empirical Complementary Cumulative Distribution Function (ECCDF) of latency of URLLC users. The figures show the latency under increasing URLLC traffic load. Both figures demonstrate the superiority of DQLD over KPPF despite that KPPF applies hard QoS rule for scheduling URLLC users first. In particular, Fig. \ref{fig:urllcLatency01} demonstrates about $8$ ms improvement at the $10^{-4}$ percentile and at $1$ Mbps offered load. Furthermore, as offered load increases, KPPF fails to maintain a reasonable performance for URLLC users. 

\begin{figure}
    \centering
    \includegraphics[scale=0.26]{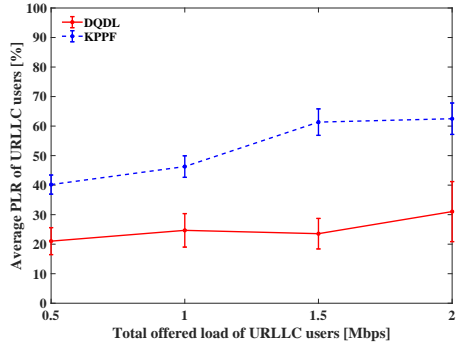}
    \caption{Packet loss rate of URLLC users versus total URLLC offered load.}
    \label{fig:urllcPlr}
\end{figure}

\begin{figure}
    \centering
    \includegraphics[scale=0.26]{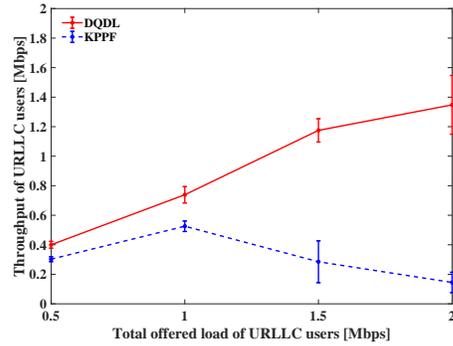}
    \caption{Sum rate of URLLC users versus total URLLC offered load.}
    \label{fig:urllcRate}
\end{figure}

\begin{figure}[t]
    \centering
    \includegraphics[scale=0.26]{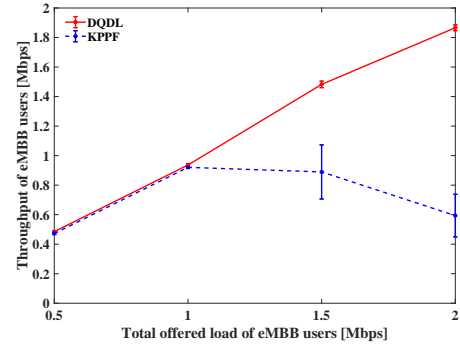}
    \caption{Sum rate of eMBB users versus total eMBB offered load.}
    \label{fig:embbRate}
\end{figure}

In Fig. \ref{fig:urllcLatency02}, the latency performance of KPPF degrades significantly, whereas DQLD was able to achieve much lower latency with about $350$ ms difference with respect to KPPF at $2$ Mbps.
The significant performance degradation of KPPF is attributed to a high Packet Loss Rate (PLR) as shown in Fig. \ref{fig:urllcPlr}. The figure presents the PLR of URLLC users under different traffic loads. As seen in Fig. \ref{fig:urllcPlr}, DQLD demonstrates a $50\%$ improvement in PLR compared to KPPF. Furthermore, Fig. \ref{fig:urllcRate} presents the achieved rate of URLLC users under different URLLC traffic load. Again, DQLD outperforms KPPF. In fact, increasing the traffic load impacts KPPF significantly, where $1$ Mbps constitutes a break point for the algorithm. It is worth mentioning that the simulation is performed by increasing traffic loads of both URLLC and eMBB users simultaneously. For example, $1$ Mbps in Fig. \ref{fig:urllcRate} refers to URLLC and eMBB loads (i.e., total load per gNB is $2$ Mbps). As such, increasing the offered URLLC and eMBB loads stresses both algorithms. Fig. \ref{fig:embbRate} presents the achieved rate of eMBB users under different traffic load. Again, the same trend appears for KPPF, where $1$ Mbps constitutes a break point in KPPF's performance, whereas DQLD demonstrates an ability to balance resources among users and satisfy the conflicting QoS requirements. 
\section{Conclusion}\label{sec:conclusion}
In this paper, we address the problem of QoS-aware clustering (for beamforming) and resource allocation in 5G millimeter wave networks. We proposed an online clustering algorithm for identifying the number and structure of beams to cover network users, in addition to an LSTM-based deep reinforcement learning to perform resource allocation within each beam. The proposed algorithm is compared to a baseline that uses K-means for clustering and priority-based proportional fairness for resource allocation.  Simulation results reveal that the proposed algorithm outperforms the baseline in terms of latency, reliability, and rate of URLLC users as well as rate of eMBB users. 
\section*{Acknowledgment}
This research is supported by U.S. National Science
Foundation (NSF) under Grant Number CNS-1647135 and Natural Sciences and Engineering Research Council of Canada (NSERC) under Canada Research Chairs Program. 

\bibliographystyle{ieeetr}
\bibliography{reference}

\end{document}